\documentclass[default,iicol]{sn-jnl}% Default with double column layout
\setcitestyle{square, comma, numbers,sort&compress, super}
\jyear{2023}%

\theoremstyle{thmstyleone}%
\theoremstyle{thmstyletwo}%

\theoremstyle{thmstylethree}%

\begin{document}
	
	\title[Chaotic diffusion in the action and frequency domains]{Chaotic diffusion in the action and frequency domains: estimate of instability times}
    \subtitle{\footnotesize Submitted to \textit{European Physical Journal Special Topics} on $15/$May$/2023$, accepted on $09/$November$/2023$.}
    
    % \titlenote{notinha do preprint}
	
	
	\author[1,2]{\fnm{Gabriel} \sur{Teixeira Guimar\~aes}}\email{gabriel.guimaraes@usp.br, gabriel.guimaraes@grad.nao.ac.jp}
	
	\author[1]{\fnm{Tatiana} \sur{Alexandrovna Michtchenko}}\email{tatiana.michtchenko@iag.usp.br}
	
	\affil[1]{\orgdiv{Instituto de Astronomia, Geofísica e Ciências Atmosféricas}, \orgname{Universidade de São Paulo}, \orgaddress{\street{Rua do Matão, 1226 - Butantã}, \city{S\~ao Paulo}, \postcode{05508-090}, \state{S\~ao Paulo}, \country{Brazil}}}
	
	\affil[2]{\orgdiv{Graduate School of Science}, \orgname{The University of Tokyo}, \orgaddress{\street{7-3-1 Hongo, Bunkyo-ku}, \city{Tokyo}, \postcode{113-0033}, \country{Japan}}}
	
	\abstract{\textbf{Purpose:} Chaotic diffusion in the non-linear systems is commonly studied in the action framework. In this paper, we show that the study in the frequency domain provides good estimates of the sizes of the chaotic regions in the phase space, also as the diffusion timescales inside these regions. 
		
		\textbf{Methods:} Applying the traditional tools, such as Poincar\'e Surfaces of Section, Lyapunov Exponents and Spectral Analysis, we characterise the phase space of the Planar Circular Restricted Three Body Problem (PCR3BP). For the purpose of comparison, the diffusion coefficients are obtained in the action domain of the problem, applying the Shannon Entropy Method (SEM), also as in the frequency domain, applying the Mean Squared Displacement (MSD) method and Laskar's Equation of Diffusion. We compare the diffusion timescales defined by the diffusion coefficients obtained to the Lyapunov times and the instability times obtained through direct numerical integrations. 
		
		\textbf{Results:} Traditional tools for detecting chaos  tend to misrepresent regimes of motion, in which either slow-diffusion or confined-diffusion processes  dominates. The SEM shows a good performance in the regions of slow chaotic diffusion, but it fails to characterise regions of strong chaotic motion. The frequency-based methods are able to precisely characterise the whole phase space  and the diffusion times obtained in the frequency domain present satisfactory agreement with direct integration instability times, both in weak and strong chaotic motion regimes. The diffusion times obtained by means of the SEM fail to match correctly  the instability times provided by numerical integrations.
		
		\textbf{Conclusion:} We conclude that the study of dynamical instabilities in the frequency domain provides reliable estimates of the diffusion timescales, and also presents a good cost-benefit in terms of computation-time.}
	
	
	\maketitle
	\section{Introduction}
	\label{sec1}
	The problem of chaotic diffusion has been studied to understand general chaotic phenomena in the non-linear dynamical systems, and, in particular, the discrepancy between Lyapunov Times ($\tau_{lyap}$) and Macroscopic Instability Times ($\tau_{inst}$, times related to collisions/ejections, close approaches and orbit-crossings) of the motion of celestial bodies (\cite{milani}).
	
	The chaotic diffusion in the action space is a commonly used approach  (\cite{froeschle2005}, \cite{cachucho}, \cite{marti2016}), however, the analysis of the diffusion in the frequency domain shows some advantages. As shown in \cite{laskar1990}, the existence of chaotic zones is much more visible in the frequency domain than in the action domain. In addition, it is also independent on the choice of the coordinates and shows sensibility to the phenomena of slow-diffusion and weak-chaos (\cite{laskar1993}).
	
	Here, we investigate the chaotic diffusion in the PCR3BP  applying both kinds of tools, estimating the  diffusion coefficients and diffusion timescales and comparing the results obtained. Finally, we compare the calculated timescales with instability times obtained through the calculation of the Lyapunov times and direct integrations of equations of motion.% \textbf{and compare how instability times obtained through distinct frameworks relate both quantitative and qualitatively to the ones obtained through direct numerical integrations}.
	
	Throughout this paper, ``diffusion times'' will refer to the times obtained by means of the calculated diffusion coefficients, while ``instability times'' are the times obtained by means of the direct numerical integrations of exact equation of motion of the PCR3BP.
	
	\section{Methods and Procedures}\label{sec2}
	
	The chaotic phenomena in the PCR3BP are analysed applying several well known methods for detection of chaotic behaviour, such as the Poincaré Surfaces of Section (S.O.S, \cite{henon}), the Fast Lyapunov Indicator (FLI, \cite{fli}) and the Spectral Number Method (SN, \cite{outer}). Each of these methods is based on the distinct theoretical background:
	
	\begin{itemize}
		
		\item {The Poincaré surfaces of section can be interpreted as a reduction of dimensionality of conservative problems with two degrees of freedom to one. Their dynamics is fully represented on the plane by the trajectories,  which are: fixed points corresponding to periodic orbits, smooth curves and chains of islands corresponding to  quasi-periodic and resonant orbits (commensurability of independent frequencies of motion), and randomly scattered points indicating the regions of chaotic motion;}
		\item {The FLI measures the maximum value of the deviation vector defined by the Jacobian of the flux corresponding to the equations of motion. It  is  related to the Lyapunov Exponents, which estimate the rate of exponential divergence of two close orbits within the phase space (the inverse of which is the Lyapunov Time $\tau_{lyap}$);}
		\item {The SN counts the number of frequency peaks, obtained by means of the Fast Fourier Transform (FFT), which are above an arbitrary defined value, introduced to discern between a numerical noise and a real signal. Its use as a indicator of chaos is based on the fact that the power spectra of regular orbits present a countable number of well defined frequencies, while the spectra of chaotic orbits are composed of a high number of frequencies, which are not clearly identified with  the proper frequencies of the problem.}
	\end{itemize}
	
	{In order to deeper understand the dynamical structure of the phase space and quantify the chaotic diffusion in the frequency and actions domains, we calculate diffusion coefficients applying The Spectral Analysis (SAM) and the Wavelet Analysis (WAM) methods, and the Shannon Entropy Method (SEM), respectively. }
	
	Finally, the estimates of diffusion timescales defined by the diffusion coefficients, are compared to instability timescales obtained through direct numerical integrations and the calculation of the Lyapunov times.
	
	\section{Diffusion Frameworks}\label{secdiff}
	\subsection{Diffusion in the action domain}\label{subsec3}
	
	Diffusive phenomena in the dynamical systems are usually studied in the action space framework, in which several  relationships between $\tau_{lyap}$ and $\tau_{inst}$ have been found (\cite{holman1997}, \cite{varvoglis}, \cite{cachucho}, \cite{batygin2015}). Nonetheless, most of analytical approaches are restricted to the specific Hamiltonian models and  the individual resonant geometry.
	
	Modern, more generic approaches, are generally based on assumption that the chaotic motion is driven by a random-walk-like dynamics related to normal diffusion (or brownian motion) and approach it by means of time-evolution of the variance of an ensemble of particles (\cite{froeschle2005}, \cite{lega2010}). Nonetheless, as shown in \cite{cordeiro2005}, there exists sub- and super-diffusive phenomena in the Solar System, which cannot be instantly linked to instability timescales.
	
	In this work, we apply the SEM, recently introduced in \cite{shannon0}, to investigate the diffusion in the action space domain. The method, which has been explored extensive and intensively (\cite{shannon1}, \cite{alves}), yields precise estimates of diffusion rates and timescales while overcoming previous methods shortcomings and being constantly enhanced. % without presuming stochasticity of distinct regions of the phase space.
	
	\subsection{Diffusion in the frequency domain}\label{subsec33}
	
	The study of chaotic diffusion in the frequency domain  is based on the well-known fact that the independent frequencies of regular orbits remain constant in time, while those of chaotic motion change during the motion (\cite{laskar1993}, \cite{wavelets}). 
	
	One notable advantage of working in frequency space is  the fact that the independent frequencies of dynamical systems are invariant under transformation of coordinates \cite{laskar1993}, making them uniquely defined regarding the coordinate system approached for the study.
	
	Besides allowing a simple detection of the chaotic zones of the Solar System (\cite{laskar1990}), the study of dynamical systems in the frequency domain makes it easier to reveal diffusive phenomena, once the web of distinct resonances existent within the problem can be easily retrieved (\cite{price2015}).
	
	In  addition, as suggested in \cite{laskar1993}, there exists a relationship between the diffusion of the frequency in the action space and the diffusion of the frequency in time, which takes the form of the Diffusion Equation:
	
	\begin{equation}
		\nabla^{2}f(x,t)\propto \frac{\partial}{\partial t} f(x,t).
		\label{nabla}
	\end{equation}
	In order to follow the original terminology used in \cite{laskar1993}, hereafter we will refer to it as $\delta\delta_{x}f(x,t)\propto \delta_{t}f(x,t)$.

	\section{Diffusion Coefficients and Diffusion Times}\label{sec3}
	
	In order to quantify chaotic diffusion, we introduce several diffusion coefficients corresponding to the different methods applied in the study of chaotic diffusion in both the action and frequency domains.
	
	\subsection{Shannon Entropy}
	
	Applied in the studies of diffusion in the action domain, SEM is based on the occupation rate $q_{0}(t)$ of cells on a grid of $q$ equally-sized cells in the action space $I, J$ (\cite{shannon1}). 
	
	The pair of actions $I(t), J(t)$ will evolve in time $t$ and occupy a number of cells. The diffusion coefficient in the action space is then defined as 
	\begin{equation}
		\label{eq:SN}
		D_{S} = \frac{\Sigma(I, J)}{q}\left\langle \frac{dq_{0}(t)}{dt}\right\rangle,
	\end{equation}
	where the term outside the brackets is an area-dependent constant and the brackets $\left\langle \dots\right\rangle$ mean the spatial averaging procedure of the particles of the ensemble.
	
	\subsection{Equation of Diffusion}
	
	From Equation (\ref{nabla}), we can immediately retrieve two frequency-based diffusion coefficients, one related to frequency diffusion in space and the other in time.
	
	Indeed, the left side of the equation, describing the frequency diffusion in action-space, is simply computed via the second derivative of the main-amplitude frequency in the Dynamical Spectra (\cite{laskar1993}, \cite{extrasolar}). It is worth noting that it is sensitive to the settings used to construct numerically the dynamical spectra, such as the total integration time $T$ and the sampling rate $\Delta t$.
	
	The right side of the equation, describing the frequency diffusion in time, has been more extensively explored. In Laskar's original paper \cite{laskar1993} and in \cite{sylvio_freq}, it was defined as $\delta_{t} = \lvert\nu^{1}-\nu^{2}\rvert$, where $\nu^{1}$ is the frequency of highest amplitude obtained from the FFT spectrum applied to the data from the first half of the total integration time and $\nu^{2}$ is the same for the second half of the total integration time.
	
	Since the FFT provides information in frequency space, but not in time, WAM is used to precisely determine the time evolution of the independent frequencies during the orbital motion. 
	
	In face of the time-and-frequency resolution, the right-hand-side of the equation of diffusion is obtained by applying the WAM to the system and, therefore, the new coefficients are defined as:
	
	\begin{itemize}
		\item $\bar{\delta}_{t}^{CWT} = \langle\partial f(t)/\partial t\rangle$ averaged time variation of the independent frequencies (\cite{arevalo}),
		\item $\delta_{t}^{CWT}=\lvert\bar{f}^{1}-\bar{f}^{2}\rvert$, that is analogous to $\delta_{t}$, but uses the mean frequencies calculated in the first and the second half of the total integration timespan.
	\end{itemize}  
	
	\subsection{Frequency MSD}
	
	In addition, the diffusion coefficients in the frequency space, which follow \cite{chirikov1979} theory of diffusion, can be obtained by means of the MSD (\cite{cincotta2000}). 
	
	We adopt the definition of the frequency diffusion from \cite{marzari_2003},  as the relative variance  $\var$ of the independent frequencies in time $D_{maz} = \var(f(t))/f(0)$, where $f(t)$ is the temporal evolution of the fundamental frequency obtained by WAM and $f(0)$ is the frequency determined at $t=0$.
	
	We also introduce a new diffusion coefficient based on the works of \cite{chirikov1979} and \cite{lega2008}; it is defined as $D_{MSD} = [f(T)-f(0)]^{2}/T$, where $f(T)$ is the frequency determined at the final moment of the numerical integration.
	
	\subsection{Dynamical Characterisation of the Phase Space of the PCR3BP}\label{sec4}
	
	The Jacobi integral of the PCR3BP is written in the rotating reference frame as 
	\begin{equation}
		C_{J} = (x^{2}+y^{2})+2\left(\frac{1-\mu}{r_{1}}+\frac{\mu}{r_{2}}\right)-(\dot{x}^{2}+\dot{y}^{2}),
		\label{const_jacobi1}
	\end{equation}
	where $(x,y)$ are the coordinates of a massless particle, moving under gravitational influence of the two massive bodies. The distances of the particle to the body of mass $1-\mu$ and the body of mass ${\mu}$ (in units of the total mass) are $r_{1}$ and $r_{2}$, respectively. 
	
	In this work, we choose $\mu=0.0009537$, whose small value places the large body of mass $1-\mu$ near the origin and the smaller body of mass $\mu$ very close to 1 on the $x$--axis (assuming the distance between two massive body as unit). $C_{J}$ is a constant in units of energy, whose value is chosen here as $C_{J} = 3.03$. More information regarding the PCR3BP, can be found in the Chapter 3 in \cite{murray}. 
	
	We numerically integrated 201 initial conditions over $T = 10^{4}$ orbital periods in the interval $0.05 \leq x(0) \leq 0.75$, with fixed $\dot{x}(0) = 0$, $y(0) = 0$ and $\dot{y}(0)$ corresponding to the chosen value of $C_J$. For all 201 particles, we calculate the FLI, the SN and the time evolution of the independent frequencies  using the WAM. An ensemble composed of 5 particles, from the vicinity of each initial condition above, was also integrated over $T = 10^{3}$ orbital periods, in order to obtain the diffusion coefficient (\ref{eq:SN}). For all particles from the ensemble, the complete spectral analysis (SA) of the orbital motion was also done.
	
\begin{figure*}[!ht]
		\resizebox{\hsize}{!}{\includegraphics{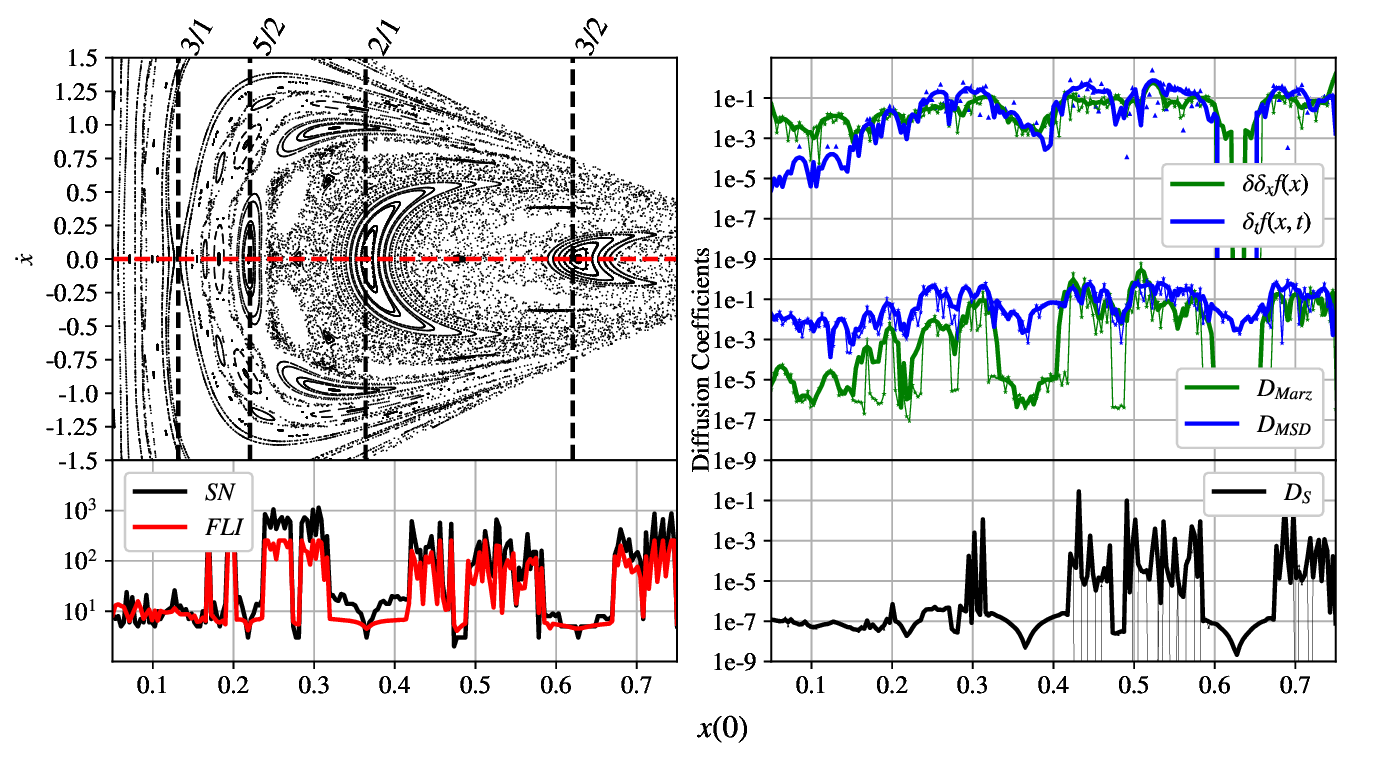}}
		\caption{\footnotesize Top left: Poincaré sections, nominal positions of the main resonances are indicated by the vertical dashed lines.\textbf{ }Bottom left: Evolution of the FLI and the SN along the $x$--axis. Top right: Diffusion Coefficients obtained using Laskar's Equation of Diffusion. Middle right: Diffusion coefficients obtained using the MSD. Bottom right: Diffusion Coefficients obtained using the SEM.}
		\label{fig:sub1}
	\end{figure*}

	The main results {obtained are shown in} Figure \ref{fig:sub1}.
	
	On the Poincaré surface of section (top-left panel) we can observe several dynamical features, mostly the resonant islands of stable motion related to the two first-order resonances, $ 2/1 $ and $ 3/2 $, with the centres at $x(0) \cong 0.37$ and $x(0) \cong 0.62$, respectively, and the third-order resonance $ 5/2 $, with the centre at $ x(0) \cong 0.22 $. The saddle-point of the strong  second-order resonance $3/1$ is around $ x(0) \cong 0.13$. The region between the low-order resonances is a sea of chaotic motion. The region close to the origin is filled by the high-order resonances of decreasing width, which are separated by the thin stochastic layers.	
	
    	Both the FLI and the SN (red and black curves on the bottom-left panel, respectively) allow us to distinguish between the distinct regimes of motion observed on the top-left panel: both are converging to the minimal values at the centres of resonances, and are suddenly increasing during the passages through the stochastic layers of the resonances.
	
	However, in the vicinity of the large body (close to the origin), the SN and FLI show more complicated behaviour, probably, due to the high density and overlap of the high-order resonances in this  region.
	
	The diffusion coefficients obtained using Laskar's relation (top-right panel) show the expected behaviour inside a region of confined chaos, where SN and FLI do not show good agreement. The coefficients obtained from the temporal part of the equation ($\delta_{t} f(x, t)$) seem to better reproduce the dynamics of the system in the region nearer the body of mass $1-\mu$ located approximately at the origin ($x(0)\approx 0$). This is due to the fact that  the spatial part of Equation (\ref{nabla}) is directly obtained from the dynamical spectra and, consequently, undergone to same limitations, such as the finite integration time $T$ and the sampling rate $\Delta t$. Thus, hereafter only temporal diffusion of fundamental frequencies will be analysed.
	
	The diffusion coefficients shown on the middle-right panel in Figure \ref{fig:sub1} are both based on the time evolution of the frequencies  and show similar behaviour. However, Marzari's method (green curve) shows more sensibility in the stable and slow-diffusion regions of the phase space, when compared to the Froeschl\'e method (blue curve). 
	
	The coefficient of the SEM (bottom-right panel) allows us to distinguish between regions of the regular and chaotic motion, being more sensitive to the regions where slow diffusion dominates, while its behaviour on the vicinity of strong chaotic regions shows sudden ``jumps'' and erratic variations, contrasting to the well behaved stable/slow-diffusing regions.

	\begin{figure*}[h!]
		\resizebox{\hsize}{!}{\includegraphics{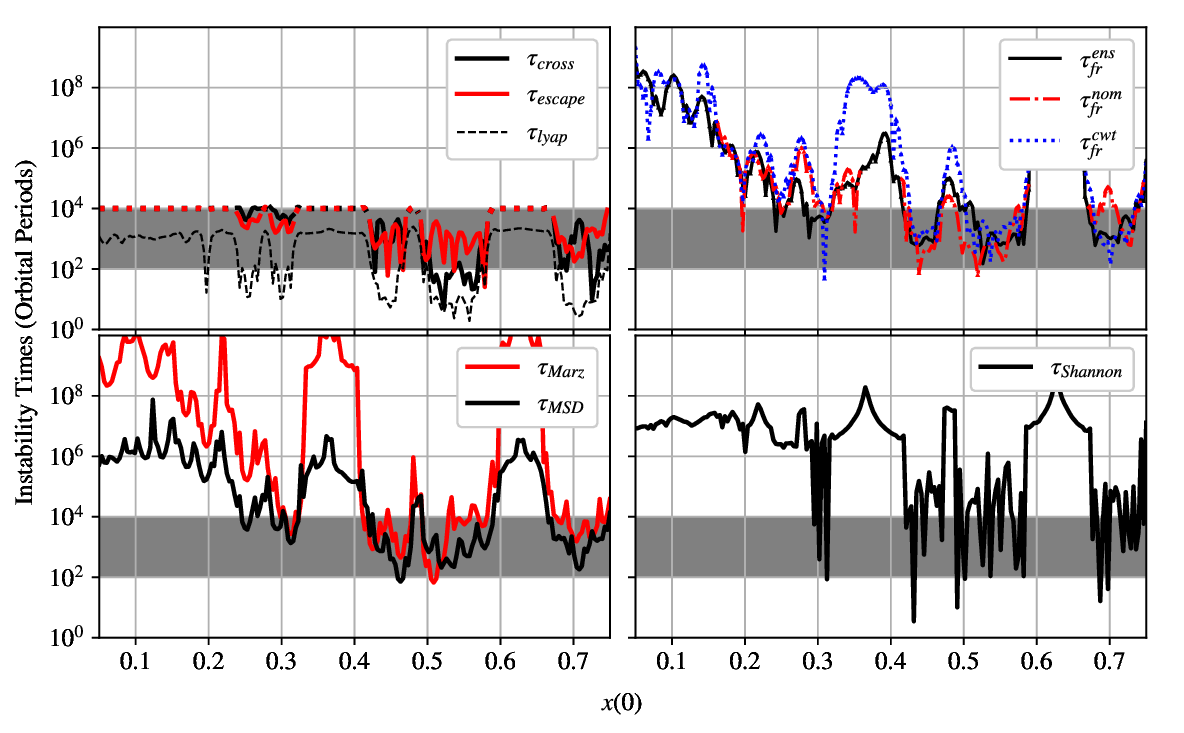}}
		\caption{\footnotesize Instability times obtained by distinct methods. Gray strips show the range of the instability times obtained through the direct integrations. Top left: Direct integrations (red curve for escape/collision times, black solid curve for orbit-crossing times and black dashed for Lyapunov times). Top right: Using \cite{robutel2006} approach, using an ensemble of particles ($\tau_{fr}^{ens}$, solid black curve), one particle ($\tau_{fr}^{nom}$, dashed red curve) and mean frequencies using WAM ($\tau_{fr}^{cwt}$, dashed blue curve). Bottom left: Using Marzari's Method ($\tau_{marz}$) and Froeschl\'e's ($\tau_{MSD}$). Bottom right: Using coefficients obtained by SEM.}
		\label{fig:sub2}
	\end{figure*}

	\subsection{Estimate of Diffusion Times}
	
	Using the values of the diffusion coefficients obtained above, we can estimate the corresponding diffusion times (\cite{froeschle2005}, \cite{robutel2006}, \cite{alves}) and compare them with instability times obtained through direct integrations. The values obtained are shown in Figure \ref{fig:sub2}. 
 
 On the top-left panel, we plot the Lyapunov time (black dashed curve), and the times obtained through the direct numerical integrations, $\tau_{cross}$ {(times needed to orbit intersections occur, in solid black lines)} and $\tau_{escape}$ {(times needed the particle either collides with one of the masses or be ejected from the system, in solid red lines)}.  In the $x(0)$--segments, where no collides and/or ejections  were detected during the integrations, the solid curves are replaced by the dotted curves.  For the purpose of comparison, the range of the typical instability times ({those in which most of the $\tau_{cross}$ and $\tau_{escape}$ obtained lie within}) is indicated by the grey strips on all panels in Figure \ref{fig:sub2}. 
	
	In the region close to the origin, the instability times are not corroborated by the results provided by the chaos indicators, once chaotic motion is confined and slowly diffusing, as shown on the previous session. Farther from the origin, the Lyapunov times are related to the escape/orbit-crossing times, following results obtained in \cite{lecar1992}, in particular, in the regions of strong chaotic motion. 
	
	The diffusion times obtained using the time-related part of Laskar's equation of diffusion $ \tau_{fr} $ \cite{robutel2006} (top-right panel in Figure \ref{fig:sub2}) mostly agree with the direct integration times. The times calculated from diffusion coefficients attained from simple FFT (\cite{laskar1993}), for short-term integrations using a single particle or the ensemble, are shown by the red and black curves, respectively. The blue dashed curve was obtained by the WAM, analysing the results of the long-term integrations. The agreement between all three approaches and the instability times shows robustness of the method regarding the number of particles and the integration times.

	The diffusion times obtained using the MSD (bottom-left panel) show variable agreement with the direct integration times: in the regions of weak chaos, the diffusion times based on Froeschl\'e's definition ($\tau_{MSD}$, black curve) shows some agreement with $\tau_{cross}$, as postulated in \cite{lecar1992}.  The agreement with $\tau_{escape}$ is better in the regions of strong chaotic motion. 
	
	The diffusion times obtained from the Marzari's coefficient ($\tau_{Marz}$, red curve) are able to put in evidence the regions of the long-term stable motion, even tough this method seems to overestimates the instability times, when compared to the MSD.
	
	The diffusion times estimated through the SEM (bottom-right panel) show certain agreement in the regions of slow diffusion and weak chaos, but fail to correctly determine diffusion timescales in regions of strong chaotic motion. This behaviour is mostly due to the fact that the system is not ergodic, but instead shows great complexity of phenomena on the vicinity of the resonances (\cite{alves}).
	
	\section{Conclusions}
	
	In this paper, we analyse the performance of the different known methods for detecting and quantifying chaotic motion, applying them to the PCR3BP. The results obtained are compared and some conclusions are done.
	
	The study of chaotic diffusion in the frequency space is shown to be of good cost-benefit, in terms of both accuracy and computation times. On the other side, the application of the SEM is proved be computationally time-consuming in our case.
	
	The frequency based methods for estimating diffusion timescales achieve better agreement with the instability times obtained through the direct numerical integrations when compared to the SEM estimates. Moreover, the SEM does not seem to yield consistent agreement with instability times in regions of strong chaotic motion when compared to estimates obtained in frequency space. 
	
	Nonetheless, it is a recent method and new properties and applications regarding the SEM are still being published. Robust mathematical development (\cite{shannon2}) and precise estimates for diffusion times (\cite{alves}) been achieved in other studies.
	
	It is important to emphasise that all the indicators here studied show the convergent results in the regions of periodic and quasi-periodic motion.  Moreover, a powerful aspect revealed by these methods consists in possibility to estimate instability times without the need to integrate for very long times, of the order of the lifetime of a system.
	
	Generally, the use of the different tools for the characterisation of dynamical systems ensures greater accuracy (\cite{cincotta2000}), however, the economy of computational resources should be kept in mind during the choice of a tool to be applied.

	Further studies must be done, in order to verify robustness of the methods, as well as the longer integration times and additional applications to the different dynamical systems; nonetheless, our results indicate the reliability of the tested tools in the frequency domain for a fast and reliable dynamical characterisation of system. Regarding the results obtained by means of the SEM, more studies should be conducted in distinct physical systems and numerical setups.
	
	Furthermore, we hope to provide a concise and simple compilation of the main methods for estimating diffusion timescales existing in the literature, so to create a reference material for those interested in dynamical characterisation of dynamical systems with a diverse set of tools and frameworks.
	
	\footnotesize{\textbf{Acknowledgements}: This work was completed with the support of a CAPES scholarship (process number 88887.496871/2020-00)}.
	
	\footnotesize{\textbf{Data availability:} No Data are associated with the manuscript.}
	
	{\small{\bibliography{sn-bibliography}}}
\end{document}